\pdfoutput=1
\documentclass[preprint,aps,showpacs,preprintnumbers,amsmath,amssymb]{revtex4}
\usepackage{graphicx}
\usepackage{dcolumn}
\usepackage{bm}
\usepackage{color}
\begin{document}
\title{Topological properties of the mesoscopic graphene plaquette:
QSHE due to spin imbalance}
\author{B. Ostahie$^{1,2}$, M. Ni\c t\u a$^{1}$ and A. Aldea$^{1}$}
\affiliation{$^1$ National Institute of Materials Physics, POB MG-7,
77125 Bucharest-Magurele, Romania \\
$^2$ Faculty  of Physics, University of Bucharest, Romania}
\date{\today}
\begin{abstract}
We study the electronic properties of the confined honeycomb lattice 
in the presence of the intrinsic spin-orbit (ISO) interaction  and  
perpendicular magnetic field, and report on uncommon aspects of the 
quantum spin Hall conductance corroborated by peculiar properties of the 
edge states.
The ISO interaction induces two specific gaps in the Hofstadter spectrum, 
namely
the 'weak' topological gap  defined by Beugeling {\it et al}
[Phys. Rev. B {\bf 86}, 075118 (2012)], and  spin-imbalanced gaps 
in the relativistic range of the energy spectrum. We analyze the evolution 
of the helical states with the magnetic field and with increasing 
Anderson disorder. The 'edge' localization 
of the spin-dependent 
states and its dependence on the disorder strength is shown.
The quantum transport, treated in the Landauer-B\"{u}ttiker formalism,
reveals interesting new plateaus of the quantum spin Hall effect (QSHE),
and also of the integer quantum Hall effect (IQHE), in the energy ranges 
corresponding to the spin-imbalanced gaps.
The properties of the spin-dependent transmittance matrix that
determine the symmetries with respect to the spin, energy and magnetic field
of the longitudinal and transverse resistance are shown.
\end{abstract}
\maketitle

\section{Introduction }
The large conceptual interest for the properties of graphene 
was  motivated first by the relativistic-like effects in the 
honeycomb structure and  the opportunity for the high temperature 
relativistic integer quantum Hall effect \cite{Novoselov}.
Next, the interest was also stimulated  by the topological 
insulating properties, based on the  helical edge states of graphene, 
that support the quantum spin Hall effect (QSHE). 
The topological phase of the graphene, predicted by Kane and Mele 
\cite{Kane}, is induced by the {\it intrinsic } spin-orbit (ISO) coupling, 
which opens a {\it topological gap} between the Dirac cones 
located at  the points $K$ and $K'$ in the Brillouin zone. 
The gap is filled with helical states stretching along the edges,
which appear in pairs and carry opposite spins in opposite directions. 
One has to remind that the helical states 
are protected against disorder by the time-reversal symmetry of 
the Hamiltonian, but they are not protected against the spin-flip 
processes involved by  the Rashba-type coupling  or
against a staggered sublattice potential; a phase diagram can be 
theoretically obtained in the space of the coupling parameters 
corresponding to the different interactions \cite{Kane-Z2}.

Because of the very small spin-orbit coupling, the QSHE could not  
be proved experimentally in graphene. The  experimental endevor 
moved toward other 2D systems which show topological properties, 
like  CdTe/HgTe/CdTe \cite{Volkov, Bernevig, Koenig}
or AlSb/InAs/GaSb/AlSb \cite{Liu,Knez}
quantum wells, and toward the 3D topological insulators \cite{Hasan,Ando}. 
The honeycomb lattice remains however under  investigation, as
optical and synthetic such lattices (where the magnetic flux and  
the spin-orbit coupling strength can be artificially tuned)
were obtained  \cite{Gibertini,Tarruell}.  
Another line of investigation consists in finding techniques for 
the enhancement of the spin-orbit coupling by introducing adatoms in 
graphene \cite{Weeks, Balakrishnan} or using other 2D materials 
like silicene \cite{Ezawa, Liu1}.

When discussing different topological systems, 
 attention should be  paid to
the existence  and behavior of the different types of edge states 
(helical or chiral), which depend on various factors as the lattice 
structure, geometry of the sample, spin-orbit interaction, presence of 
the magnetic field. We remind, for instance, that even for  vanishing 
spin-orbit interaction, edge states are supported by the  zig-zag graphene  
ribbon, but not by the arm-chair ribbon. This was proved by solving 
the Dirac equation with proper boundary conditions \cite{Brey}
or by calculating the Zak invariant \cite{Montambaux}. However, when the 
intrinsic spin-orbit coupling is considered, the edge states, which 
become now spin-polarized and helical, are present both in the zig-zag 
\cite{Kane}  and arm-chair \cite{Shevtsov}  ribbons.
The relevance of the sample geometry can be  noticed also  by  unfolding 
the ribbon and imposing  everywhere vanishing boundary conditions.
For the finite-size plaquette, we find that the helical 
edge states extend all around the perimeter, looking however
different along the zig-zag  and arm-chair margin, respectively (see Fig.2).

When the system is subject to a magnetic field, which breaks the 
time-reversal symmetry, we expect interesting peculiarities of the
edge states under the mixed effect of the magnetic field and spin-orbit (SO)
interaction. Even for the torus geometry (i.e., with periodic boundary 
conditions along the both directions), when the edge states are missing,
the Hofstadter energy spectrum  exhibits relevant aspects  in the
simultaneous presence of the intrinsic SO coupling, Rashba-type SO interaction
and perpendicular magnetic field \cite{Beugeling}.
It turns out that the topological gap opened around $E=0$ \cite{note}
closes with  increasing magnetic flux, and it is {\it weak} in the 
sense that it is annihilated  by the Rashba coupling.
It was also found in the graphene ribbon subjected to a magnetic field
that an additional staggered potential \cite{note1} induces
spin imbalanced  regions in the spectrum, where the number of spin-up and
spin-down states are different \cite{Beugeling}.
For the confined graphene system, a spin imbalance will be detected
in this paper as due to the splitting induced by the ISO coupling, 
and interesting 
consequences  for the charge and spin transport will be  put forward.

In this paper we study  the confined honeycomb lattice obtained by 
imposing vanishing boundary conditions all along the perimeter. 
This  approach simulates the mesoscopic case, provides some specific 
new properties, and allows for  the calculation of the transport properties 
and  disorder effects. 
The plaquette exhibits both zig-zag and arm-chair boundaries as in Fig.1, 
and the first question concerns the fate of the helical states familiar  
from the cylinder (ribbon) geometry. 
The combined effect of the intrinsic SO coupling and perpendicular 
magnetic field on the spectral properties of the graphene plaquette 
are discussed in the next section. Some specific spectral properties 
anticipate new aspects of the charge and spin transport, which are 
presented in section III. 
The robustness of the spectral properties against the Anderson disorder 
is analyzed in a subsection. The disordered spectrum 
corresponding to helical states
exhibits a tulip-like picture 
due to existence in the graphene spectrum of some highly degenerated energies
(corresponding to the saddle points in the infinite model).
In section III we show how
the symmetry properties of the spin-dependent
electron transmittance give rise to particular features of the charge
and spin  currents, which are calculated in the 
Landauer-B\"{u}ettiker formalism 
for a four-lead device.
Both the spin and charge Hall 
conductance  exhibit supplementary plateaus corresponding  to gaps 
characterized by the  imbalance between the edge states with opposite spins.
The conclusions are summarized in the last section.

\section{Spectral properties of the  topological insulating 
graphene plaquette in magnetic field}
In this section we reveal new spectral properties of topological graphene
in perpendicular magnetic field, insisting on the features of the different 
types of edge states that result by imposing vanishing boundary conditions 
all around the perimeter of the plaquette. The localization of the wave
function and the robustness against disorder are discussed. 

We remind that the Hofstadter spectrum of the  graphene sheet 
in the absence of the spin-orbit coupling looks like a double butterfly 
\cite{Rammal, Cuniberti}, and exhibits both relativistic 
Dirac-Landau bands in the middle and conventional Bloch-Landau bands at 
the extremities of the spectrum \cite{note2}, separated by well-defined gaps.
In the case of the {\it finite}  plaquette, the  vanishing boundary 
conditions  and the perpendicular magnetic field generate {\it chiral} 
edge states that fill the gaps. The sign of chirality is determined by 
the direction  of the magnetic field, and one has to mention that the 
relativistic  and conventional edge states show opposite chirality.
A second class of edge states in the system are the {\it helical} ones, 
which appear in the presence of the ISO coupling, and are located in the 
topological gap opened by this interaction.

Our aim in this section is: i) to  note the evolution of the helical 
states with the magnetic field, ii) to evaluate the degree of  localization  
along  the edges of the helical and chiral states, iii) to identify  domains 
of imbalance between the densities of spin-up and  spin-down edge states
(where the charge and spin currents should become anomalous), 
iv) to see the effect of the Anderson disorder on  the energy spectrum and
on the 'edge' localization of helical states.  

Adopting the tight-binding representation, as the 2D honeycomb lattice 
contains  two atoms $A$ and $B$ per unit cell, we define  corresponding 
creation and annihilation operators $a^{\dagger}_{\sigma,nm},
b^{\dagger}_{\sigma,nm},a_{\sigma,nm},b_{\sigma,nm}$, 
where $\sigma=\pm 1$ is the spin index and  $\{n,m\}$ are the 
cell indices (see Fig.1). 
The Hamiltonian defined on the  honeycomb lattice can be written as:
\begin{equation}
H=\sum_{\sigma}H_0^{\sigma} + \sum_{\sigma}H_{SO}^{\sigma}~,
\end{equation}
where the first term describes the tunneling between the  nearest
neighbors, while the second one represents the intrinsic spin-orbit 
interaction.  
In the presence of a perpendicular magnetic field, described by the 
vector potential $\vec{A}=(-By,0,0)$, the first term reads:
\begin{eqnarray}
H_{0}^{\sigma}=\sum_{nm} E_a a^{\dagger}_{\sigma,nm}a_{\sigma,nm}+
E_b b^{\dagger}_{\sigma,nm}b_{\sigma,nm} 
+t(e^{i\phi(m)}a^{\dagger}_{\sigma,nm}b_{\sigma,nm} \nonumber\\ 
+ e^{i\phi(m)}b^{\dagger}_{\sigma,n+1,m}a_{\sigma,nm} 
+b^{\dagger}_{\sigma,n,m+1}a_{\sigma,nm}  
+H.c.).
\end{eqnarray}
$E_{a,b}$ are the atomic energies, $t$ is the hopping integral
between the sites $A$ and $B$, and the Peierls phase due to the 
magnetic field equals 
$\phi(m)=\pi\big(m+\frac{1}{6}\big)\Phi$, where the magnetic flux 
through the unit cell $\Phi$ is expressed in quantum flux units 
$\Phi_0=h/e$.

The intrinsic spin-orbit Hamiltonian \cite{Kane}  conserves the
electron spin $S_z$, and invokes the hopping to the six 
next-nearest-neighbors, keeping also in mind the chirality of the 
trajectory between the two sites.
In the presence of the  magnetic field,  the hopping terms acquire
a supplementary phase, and the Hamiltonian can be written in a compact
form as \cite{Shevtsov}:
\begin{equation}
H_{SO}^{\sigma}=i\lambda_{SO}\frac{1}{2}\sigma \sum_{<<nm,n'm'>>}
\nu_{nm}e^{i\phi_{nm}^{a}}~
 a^{\dagger}_{\sigma,n'm'}
a_{\sigma,nm} + (a\rightarrow b)  + H.c.~ ,
\end{equation}
where $\lambda_{SO}$ is the spin-orbit coupling constant, $\nu_{nm}=\pm 1$ expresses the  clock- or anticlockwise
chirality of the trajectory between the next-nearest-neighbors, 
and the phases $\phi_{nm}^{a},\phi_{nm}^{b}$ should be calculated 
by the integration of the vector potential along each trajectory.
The Hamiltonian (3) contains many terms and, for the reader's sake, 
we write it in detail, and show also  the  illustrative Fig.1 :

\begin{eqnarray}
\hskip-2cm
H_{SO}^{\uparrow}=i\lambda_{SO}\frac{1}{2}\sum_{nm} 
e^{i\phi_1^{a}(m)}a^{\dagger}_{\uparrow,n,m+1}a_{\uparrow,n,m}
+e^{i\phi_2^{a}(m)}a^{\dagger}_{\uparrow,n+1,m-1}a_{\uparrow,n,m}
+e^{i\phi_3^{a}(m)}a^{\dagger}_{\uparrow,n-1,m}a_{\uparrow,n,m} \nonumber\\
+e^{i\phi_1^b(m)}b^{\dagger}_{\uparrow,n+1,m}b_{\uparrow,n,m}
+e^{i\phi_2^b(m)}b^{\dagger}_{\uparrow,n-1,m+1}b_{\uparrow,n,m}
+e^{i\phi_3^b(m)}b^{\dagger}_{\uparrow,n,m-1}b_{\uparrow,n,m}~+H.c.~, 
\nonumber\\
H_{SO}^{\downarrow}=-i\lambda_{SO}\frac{1}{2}\sum_{nm}
e^{-i\phi_1^{a}(m)}a^{\dagger}_{\downarrow,n-1,m+1}a_{\downarrow,n,m}
+e^{-i\phi_2^{a}(m)}a^{\dagger}_{\downarrow,n,m-1}a_{\downarrow,n,m}
+e^{-i\phi_3^{a}(m)}a^{\dagger}_{\downarrow,n+1,m}a_{\downarrow,n,m}\nonumber\\
+e^{-i\phi_1^{b}(m)}b^{\dagger}_{\downarrow,n-1,m}b_{\downarrow,n,m}
+e^{-i\phi_2^{b}(m)}b^{\dagger}_{\downarrow,n+1,m-1}b_{\downarrow,n,m} 
+e^{-i\phi_3^{b}(m)}b^{\dagger}_{\downarrow,n,m+1}b_{\downarrow,n,m}~+ H.c..~~
\end{eqnarray}
The phases in the above equation are the following:
\begin{eqnarray}
\phi_1^a(m)&=&~\pi(m+\frac{5}{6})\Phi, ~
\phi_2^a(m)=\pi(m-\frac{1}{6})\Phi,~~~
\phi_3^a(m)=-2\pi(m+\frac{1}{3})\Phi,~~ \nonumber\\
\phi_1^b(m)&=&2\pi m \Phi,~~~~~~~~~
\phi_2^b(m)=-\pi(m+\frac{1}{2})\Phi,~
\phi_3^b(m)=-\pi(m-\frac{1}{2})\Phi~.
\end{eqnarray}
\begin{figure}[htb]
        \includegraphics[angle=-0,width=0.5\textwidth]{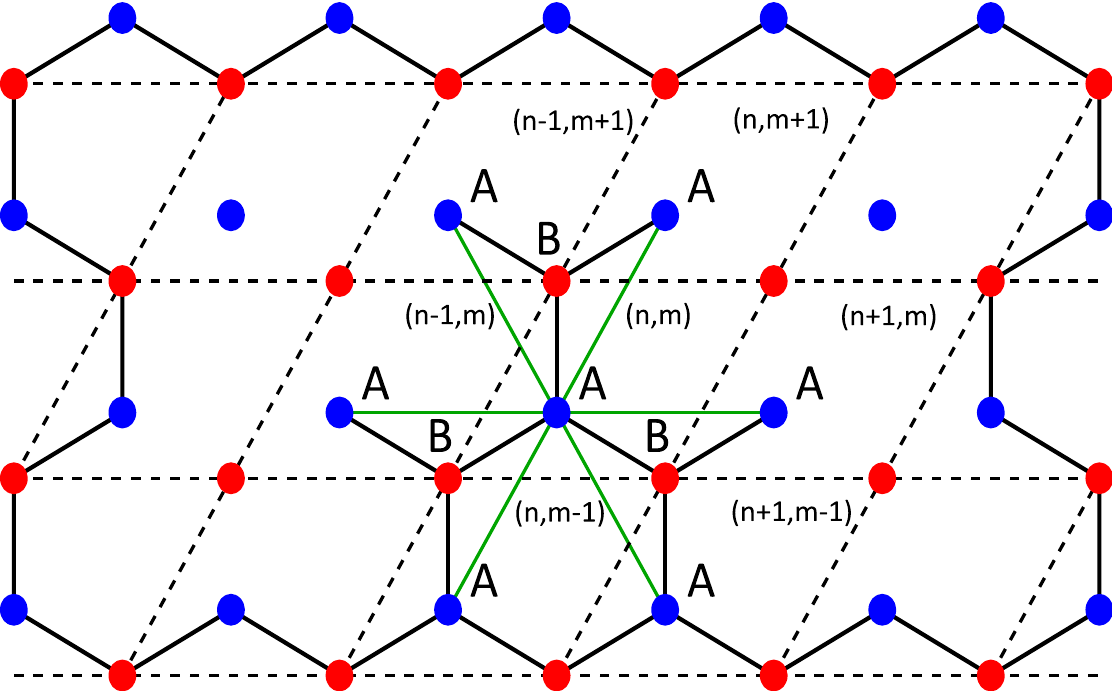}
\caption{(Color online) The sketch of the graphene plaquette 
with horizontal zig-zag and vertical arm-chair edges. 
The two type of atoms in the unit cell are A (blue) and B (red); 
(n,m) are the cell indices. The green lines connect an atom A to the six
next-nearest neighbors, while the nearest neighbors are connected 
by black lines;  the units cells are drawn with dashed lines.
The number of lattice sites is $11 \times 4$.}
\end{figure}

\begin{figure}[htb]
\includegraphics[angle=-0,width=0.5\textwidth]{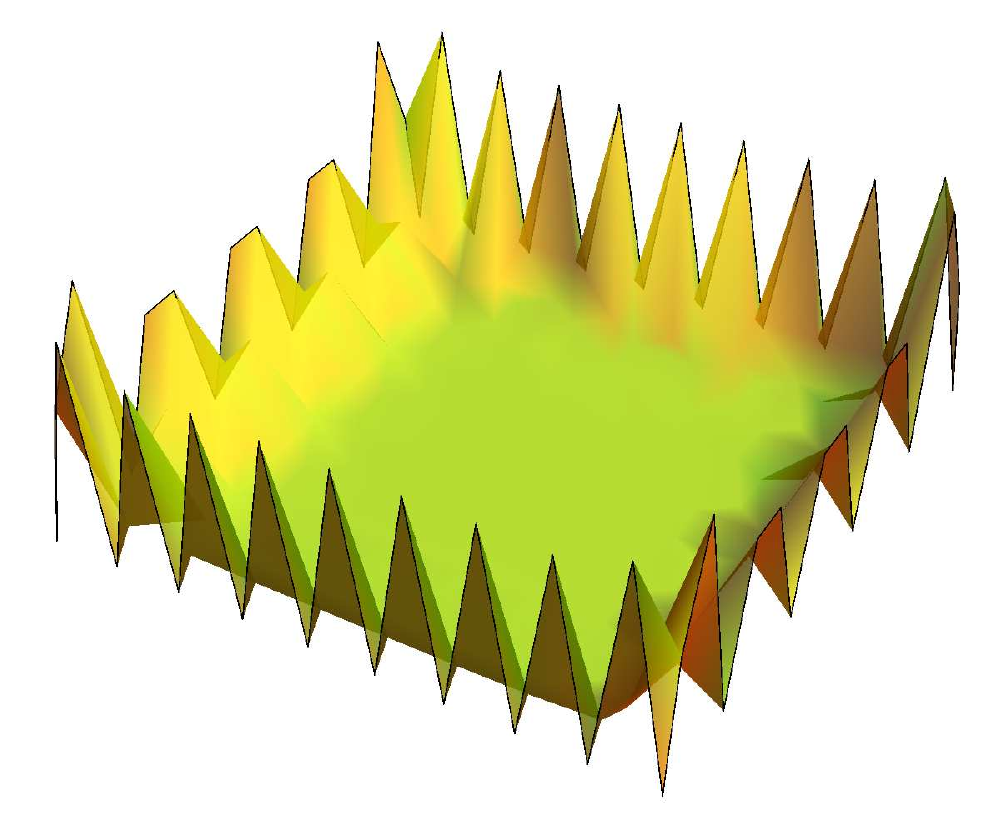}
\caption{ $|\Psi(r)|^2$ for a helical state on the graphene 
plaquette at $\Phi=0$ and $\lambda_{SO}=0.2$; a different aspect along the 
two different edges 
(zig-zag and arm-chair) is noticed. The number of lattice sites is 
$19 \times 10$.}
\end{figure}

It is worth to note  some symmetry properties of the energy spectrum.
Since the Hamiltonian (1) commutes with $S_z$, its  spectrum is
the union of the spin-up and spin-down eigenvalues $\{E_i\}=
\{E_n^{\uparrow}\} \cup \{E_n^{\downarrow}\}$, where $n=1,..,N$ 
($N$ being the total number of sites on the finite lattice).
Let $n=1$ be the index of the lowest eigenvalue for both spin-up and 
spin-down subsets. With this notation,
the symmetry of the Hamiltonian (1) generates the property 
$E_n^{\uparrow}(\Phi)=-E_{N+1-n}^{\downarrow}(\Phi)$. In words, 
this means that if  the energy $E$ belongs to the spin-up subset of the
spectrum, the energy $-E$ exists also in the spectrum, but belongs to
the spin-down subset.
One has also to note that the usual periodicity with the magnetic flux 
$E_i(\Phi)= E_i(\Phi+\Phi_0)$, which is valid at $\lambda_{SO}=0$, 
is replaced by  $E_i(\Phi)= E_i(\Phi+6\Phi_0)$ in the case of non-vanishing 
spin-orbit coupling \cite{Beugeling}.

An eigenfunction of the Hamiltonian (1) with $\Phi=0$ corresponding to a
helical edge state is shown in Fig.2. It is to observe that the state 
stretches along the whole perimeter of the plaquette, but the aspect along
the zig-zag edges differs from  that one along the arm-chair edges.

\subsection{Edge states in the 'weak' topological gap }
For vanishing  SO coupling,  the low flux range  of the Hofstadter 
butterfly of the finite-size graphene plaquette shows a thin, 
quasi-degenerate band at $E=0$, as it can be noticed in Fig.3(left).
These states correspond to the Landau band indexed by $n=0$ in the 
periodic geometry, and their number depends on the dimension of 
the plaquette. The significant changes that appear when the ISO coupling is 
introduced are only partially studied  in the presence of the 
magnetic field. One knows that the topological gap existing at 
$\Phi=0$ persists at low flux, but closes with increasing $\Phi$. 
This gap is called 'weak' in \cite{Beugeling}, and we keep the terminology.
However, the  origin and properties of the edge states filling 
the weak topological gap of the mesoscopic grahene plaquette 
have not been studied yet. They result from the simultaneous presence 
of the magnetic field and  ISO interaction, and have to justify 
the survival  of the QSHE  at non-vanishing magnetic field (see Fig.12).

\begin{figure}[htb]
\hskip-1.5cm
\includegraphics[angle=-0,width=0.9\textwidth]{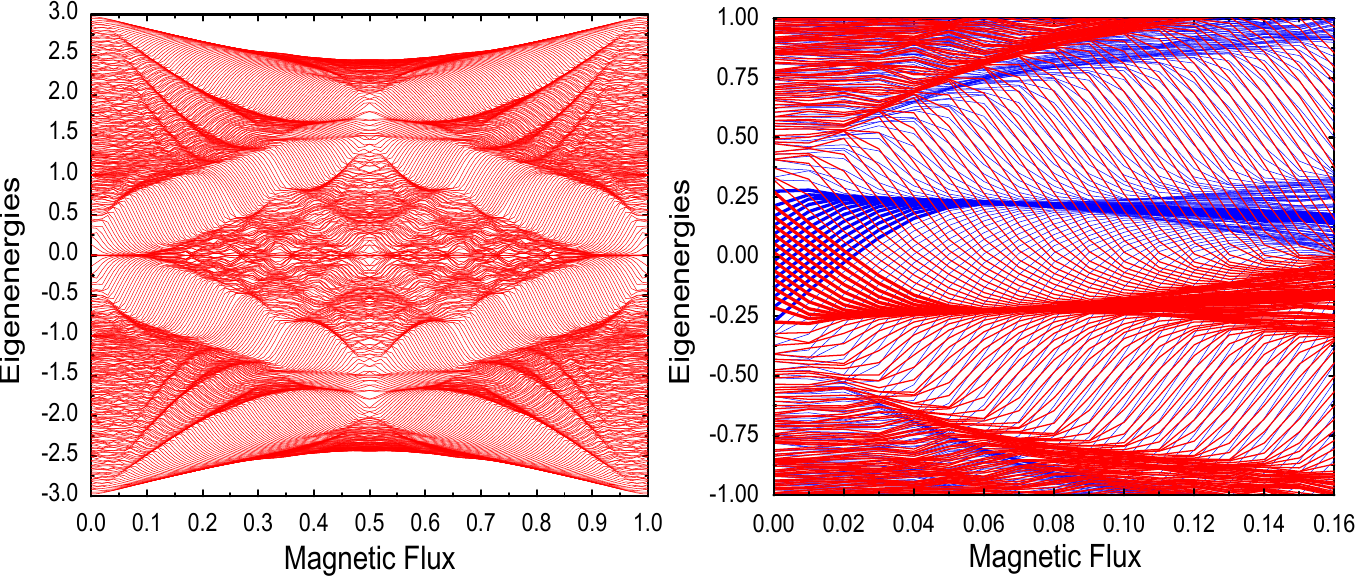}
\caption{(Color online)(left) The Hofstadter spectrum of the 
finite graphene lattice in the absence of the SO coupling; due to 
confinement, the gaps are filled with edge states.
(right) The central part of the energy spectrum showing the weak 
topological gap and two adjacent relativistic gaps in the presence 
of the ISO coupling. The spin-up eigenvalues  are colored in red, 
while the spin-down in blue. The energy is measured in units of 
hopping integral $t$, the magnetic flux in flux quanta $\Phi_0$, 
and $\lambda_{SO}=0.05$. The number of lattice sites  is $21 \times 20$.}
\end{figure}

The analysis of the edge states located in the topological gap 
will be done by  inspection of Fig.3(right) and Fig.4. 
One may identify a first class of states resulting from the splitting 
in magnetic field of the doubly-degenerated helical states existing at 
$\Phi=0$. These states are drawn in Fig.3(right) with thicker lines. 
One notices that at low flux the splitting that separates the spin-up 
and the spin-down levels increases linearly with $\Phi$, however,
at some higher magnetic flux, all these states merge into bands
that border the weak topological gap (colored in red for spin-up and 
blue for spin-down). Since $dE^{\uparrow}/d\Phi$ and 
$dE^{\downarrow}/d\Phi$ show opposite signs, the states of opposite 
spins continue to carry opposite spin currents. 
Besides, the weak topological gap  accommodates also a second category 
of edge states, which are chiral states  stemming from the adjacent 
relativistic gaps. We can see, for instance, that spin-down 
states (in blue) coming  from the relativistic gap below cross 
the red band (composed of spin-up states), enter the weak topological gap, 
and eventually merge the blue band that border the topological gap 
from above (and similarly for the spin-up red lines entering the 
topological gap  from above). 

The both types of  edge states filling the weak topological gap at
$\Phi\ne 0$, although of different provenance (helical or chiral),
show opposite  currents for opposite spins, so that the QSHE survives 
at any magnetic field as far as the  gap remains open. 
\begin{figure}[htb]
\includegraphics[angle=-0,width=0.45\textwidth]{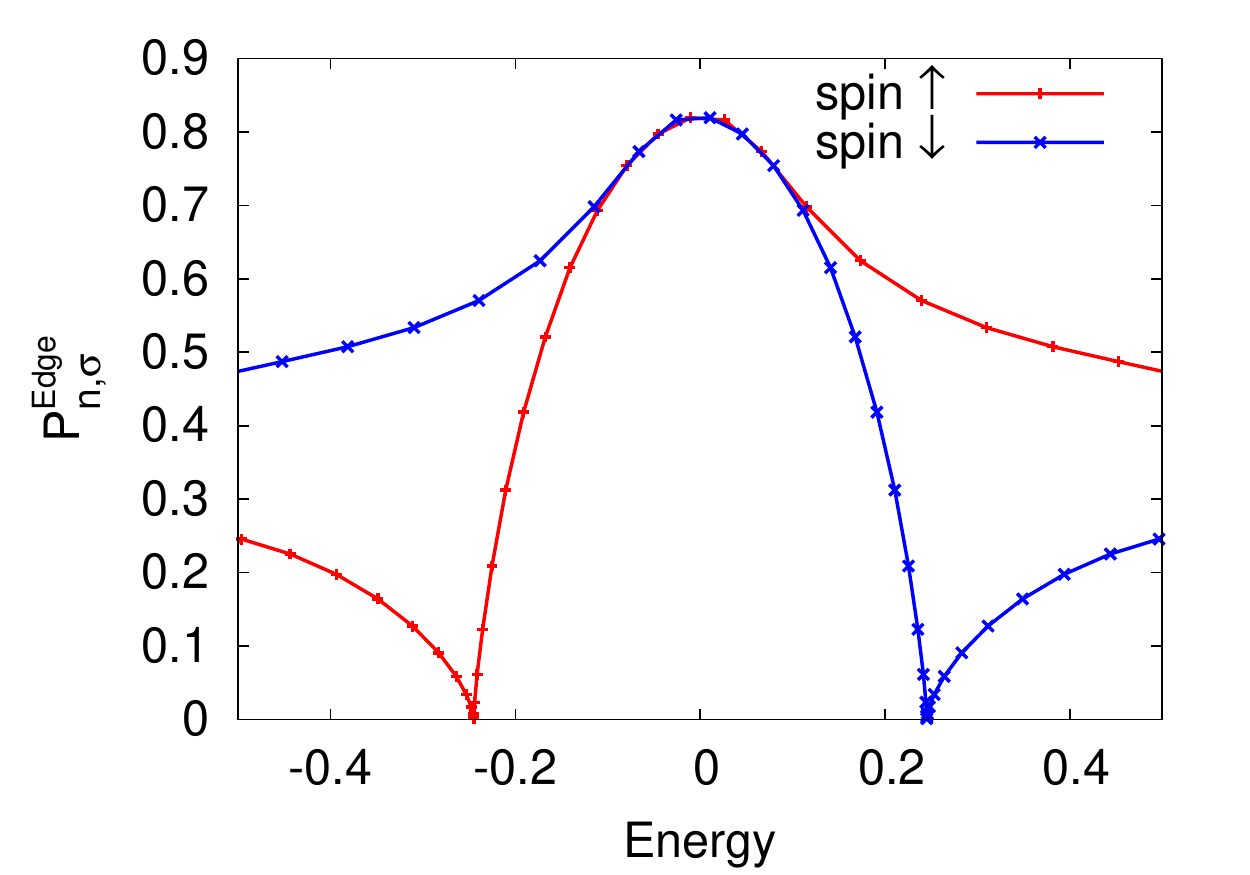}
\caption{Edge localization $P^{Edge}_{n,\sigma}$ of the spin-dependent 
eigenenergies in the range of topological and relativistic gaps. 
($\Phi=0.03\Phi_0$, $\lambda_{SO}=0.05$, the number of lattice sites 
is  $35 \times 20$).} 
\end{figure}

In what follows we examine the degree of localization of different 
edge states. The information regarding the edge localization can be 
obtained from the quantity: 
\begin{equation}
P^{Edge}_{n,\sigma}= \sum_{i\in Edge}|\Psi_{n,\sigma}(i)|^2,
\end{equation}
where the sum is taken over all sites $i$ that belong to the plaquette 
boundary \cite{Nita}. The data in Fig.4, calculated at $\Phi=0.03$,
indicate that the  states that are  close to $E=0$ are strongly localized 
along the edges. This is expected, but it is less expected that the helical 
states that converge toward the bands confining the weak topological 
gap at $E\approx\pm0.25$ are pushed away from edges, such that 
$P^{edge}$ eventually vanishes at the respective energies. 
This denotes that, while evolving into the two bands, the helical states
lose their localized character and become more similar to bulk states.

The same Fig.4 shows that outside the topological gap,
in the relativistic gaps where all states are of chiral-type, 
the edge localization depends significantly on the spin orientation.
At the same time, from Fig.3(right) one can see that the derivative 
$dE_n^{\sigma}/d\Phi$ 
is also spin-dependent.
At $E>0.25$, for instance, both  the edge localization 
and the magnetic moment of the spin-up states are higher then for the
opposite spin.

\subsection{Spin-orbit effects on the properties of the relativistic gaps}
A small  spin-orbit coupling (meaning $\lambda_{SO} \ll t$) affects 
visible only the center of the spectrum occupied by the topological gap  
and  relativistic bands and gaps. 
The extremities of the spectrum, corresponding to  
the conventional Landau bands/gaps, being  less sensitive to the 
spin-orbit coupling.  This statement is proved by Fig.5, which shows  
eigenvalues $E_n^{\uparrow},E_n^{\downarrow}$ and their corresponding index $n$, 
at a given flux.
The (quasi) horizontal lines correspond to the energy gaps 
(where the edge states are rare) and the steps correspond to the 
bands (where the bulk states are dense).
The difference between the two lines corresponding to  opposite spins 
is well visible in the energy range [-1,1], while for energies outside this
range
the  lines overlap, meaning an indistinguishable spin-orbit splitting.

\begin{figure}[htb]
\includegraphics[angle=-00,width=0.5\textwidth]{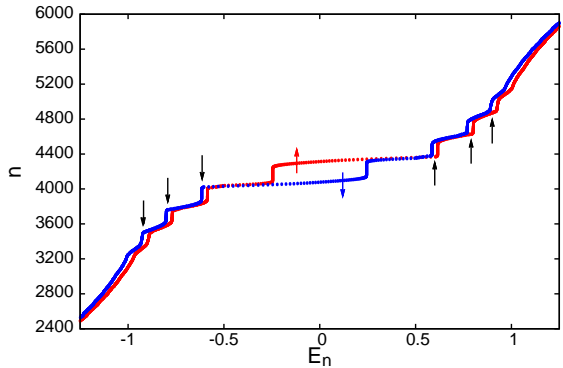}
\caption{ The eigenvalue spectrum $\{E_n^{\sigma}\}$ 
of the Hamiltonian (1) with $\Phi=0.03\Phi_0$ and $\lambda_{SO}=0.05$ 
for the finite honeycomb lattice with $105 \times 40$ sites. The spin-up energies
 are in shown in red, the  others in blue. The black arrows indicate the 
presence of the spin imbalanced gaps.}
\end{figure}

We have to  make two observations concerning the behavior of the 
spin-dependent edge states in the relativistic gaps: 
i) in contradistinction to the case of the topological gap 
(described in the previous subsection) the chirality 
$dE_n^{\sigma}/d\Phi$ of  edge states  shows now the same sign, 
independently of the spin orientation. A difference appears however 
in what concerns the magnitude of the derivative, which again is
more pronounced for the internal gaps and less evident at higher energies.
ii) The Hofstadter butterfly  exhibits the splitting of each relativistic 
band in two spin-dependent subbands. The small spin-orbit gap created 
inbetween  is filled with edge states of both spin, however, essentially, 
the number of spin-up  states differs from the number of states with  
spin-down. This denotes the existence in the energy spectrum of 
'spin-imbalanced' gaps induced by the ISO coupling \cite{note3}.
This finding should not be overlooked as it is  associated obviously
with an imbalance of the  spin currents, which may  account for  
a non-zero QSHE in  the corresponding energy range. The explicit 
calculation in the next section of the  spin-dependent electron 
transmittance  confirms this prediction. 

\begin{figure}[htb]
\includegraphics[angle=-0,width=0.5\textwidth]{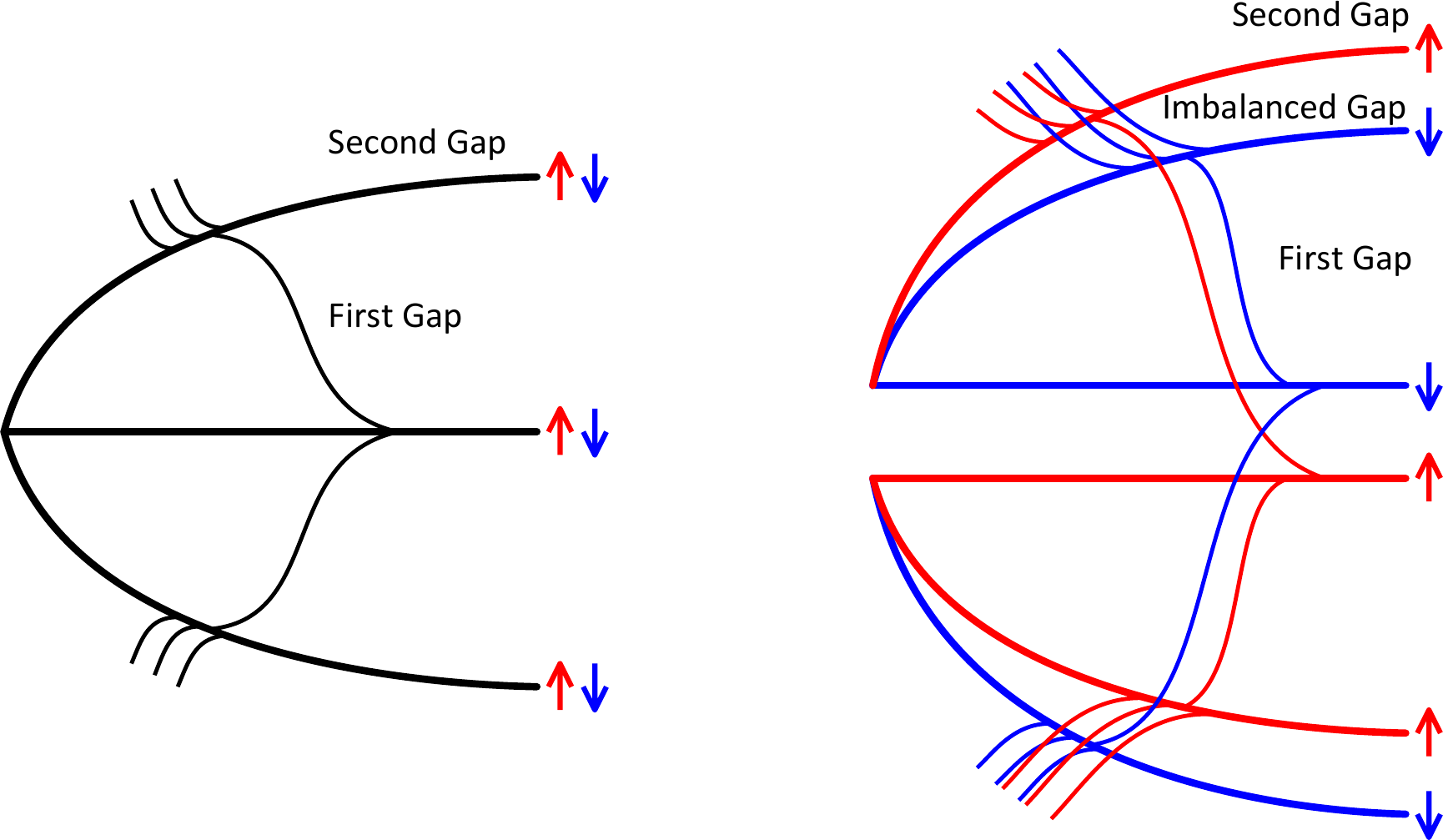}
\caption{(Color online) Schematic representation of the 
Dirac-Landau bands as function of the magnetic field: (left) in the
absence of ISO coupling the bands are spin-degenerate; (right) in the presence of 
ISO coupling each band splits into subbands of opposite spin. The gap 
created in between contains a different number of spin-up and spin-down 
states, as explained in the text.} 
\end{figure}

The sketch in Fig.6  describes the manner in which the imbalanced 
gap arises. On the left, we show the first two relativistic gaps,
separated by the relativistic band, in the case of vanishing ISO coupling, 
when all states are spin-degenerate. 
It is known  that  the number of edge states crossing the Fermi level 
at a given flux in the first relativistic gap  is 
$N_{\uparrow}+N_{\downarrow}=2$ \cite{Peres}, 
while in the second gap the number is 6.
In the right panel,  the degeneracy of the band and of the edge states is 
lifted in the the presence of the ISO coupling, and a  small spin-orbit 
gap arises between the spin-down (blue) and spin-up (red) subbands.
Now, we are interested in the  number and the spin of the  edge states 
occurring in this gap.
To this aim, considering, for instance, the upper half of the spectrum,
let us notice that the spin-up (red) edge state crosses the spin-down
subband (blue) and enter the spin-orbit gap. (At the same time, 
the spin-down edge state is absorbed in the subband of the same spin.) 
Next, we notice that three spin-down edge states originating from the 
blue subband emerges in the spin-orbit gap, then cross the subband of 
opposite spin (red), and eventually enter the second relativistic gap.
Altogether, it turns out that there are four edge states in the gap 
we look at, namely $N_{\uparrow}=1$ and $N_{\downarrow}=3$, 
fact that justifies the  term of 'spin-imbalanced' gap (see also the 
note \cite{note4}).

\subsection{Disorder effects }
Both the  helical and chiral edge states are topological states which
are robust with respect to disorder, being  protected however by 
different symmetries. 
The helical states are protected by the time-reversal symmetry 
(TRS) which is preserved by the ISO interaction, while the chiral 
edge states, despite the TRS breaking, are robust against disorder 
due to the strong magnetic field that imposes the chiral motion and 
impedes the  backscattering.
So, it is pertinent to ask whether the two types of states are equally 
robust.  Up till now, we could not give a definite answer to this question, 
and here we restrict ourselves to follow the evolution with the disorder 
of the spectral properties only at $\Phi=0$.
We use the Anderson disorder model characterized  by the  parameter 
$W$ defining the width of the diagonal disorder.  

\begin{figure}[htb]
\includegraphics[angle=-0,width=0.45\textwidth]{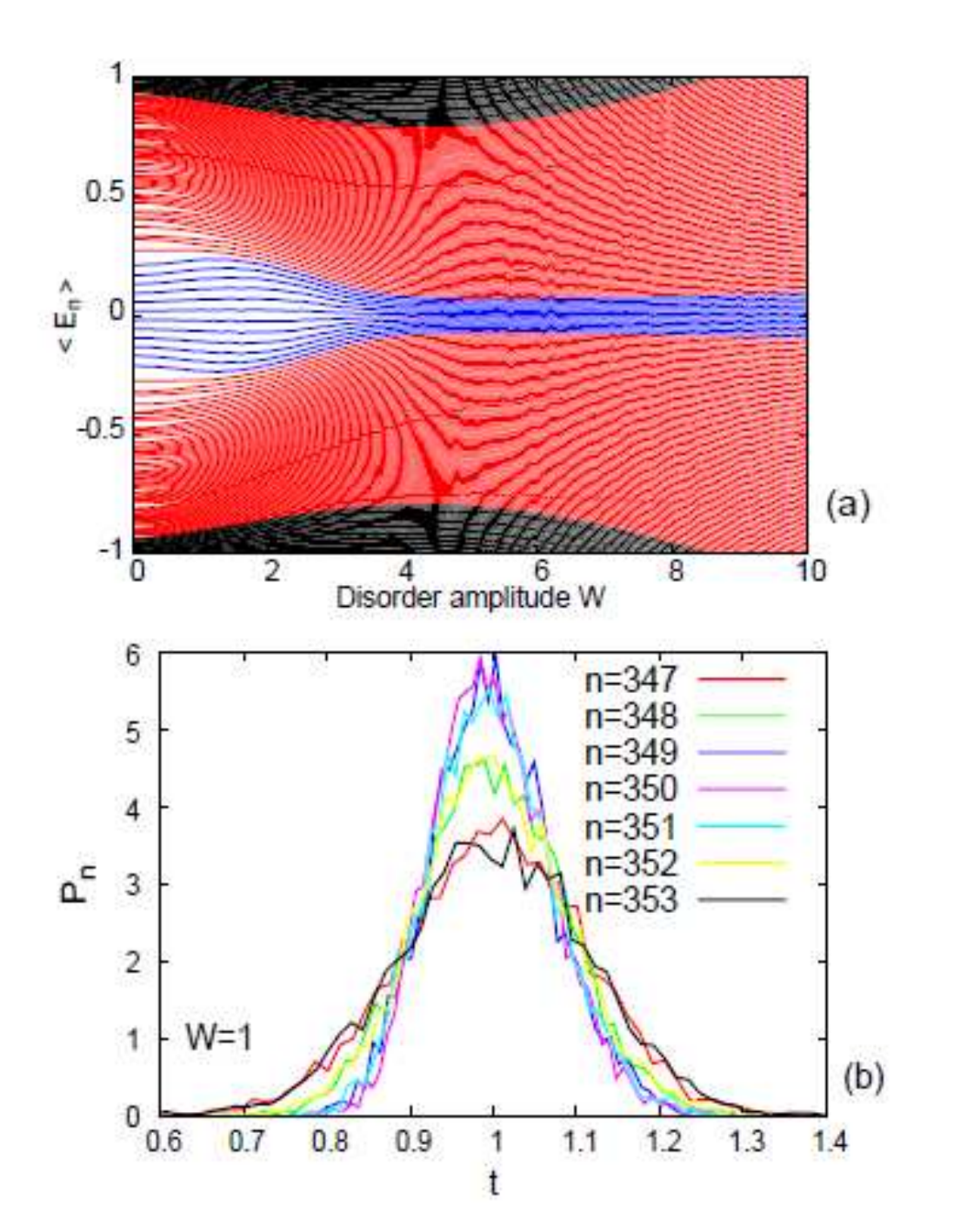}
\vskip-1cm
\caption{(Color online) (a) Disorder averaged eigenenergies $<E_n>$ 
vs. the Anderson disorder amplitude $W$ at $\Phi=0$. The topological
states (indexed by $n \in [345,356]$ are colored in blue and develop a 
{\it tulip}-like shape with increasing disorder. 
 (b) The level spacing distribution $P_n(t)$ for the helical states with
$n=347,...,353$ at $W=1$. The distribution functions are well 
fitted by  Gaussian functions; note that the states in the middle of the 
topological gap show a narrower Gaussian 
($\lambda_{SO}=0.05$, number of lattice sites is $35 \times 20$, 
number of disorder configurations is 880 (a) and 5000 (b)).}
\end{figure}

The general aspect of the disordered spectrum of the confined graphene 
lattice shown in Fig.7a is determined by the existence of regions 
that respond differently to the increase of the disorder strength. 
One knows that, at low disorder, the topological gap is not affected, 
but, on the other hand, the energy ranges with very high density of 
states about $E=\pm 1$ (which correspond in the periodic  model  to 
the saddle points $M$ in the Brillouin zone) are very sensitive to 
any disorder. The consequence is a specific {\it tulip}-like shape of the 
spectrum in the topological range $E\in [-0.25,0.25]$, depicted in 
blue in Fig.7a. 
The qualitative explanation of this shape is the following: the disordered
potential broadens the very dense spectrum close to $E= \pm 1$, 
where the  level spacing increases with the disorder strength $W$ and 
produces a 'compression' on the topological levels located in the middle of 
the spectrum. Since, according to the von Neumann-Wigner theorem 
\cite{NW,Demkov}, the energy levels cannot cross each other, 
the result is the  tulip shape of the levels in the topological range.  

\begin{figure}[htb]
\hskip-1.0cm
\includegraphics[angle=-0,width=0.45\textwidth]{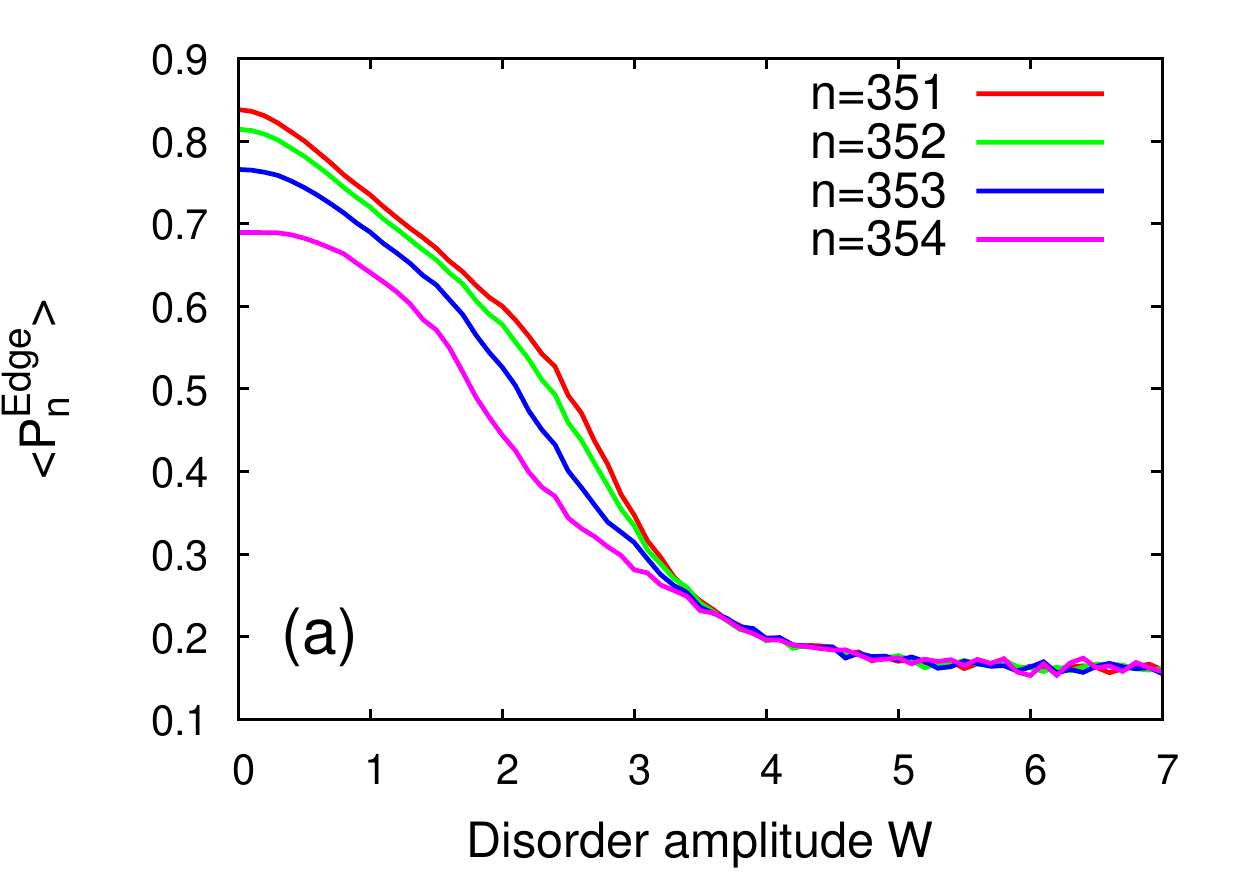}
\includegraphics[angle=-0,width=0.5\textwidth]{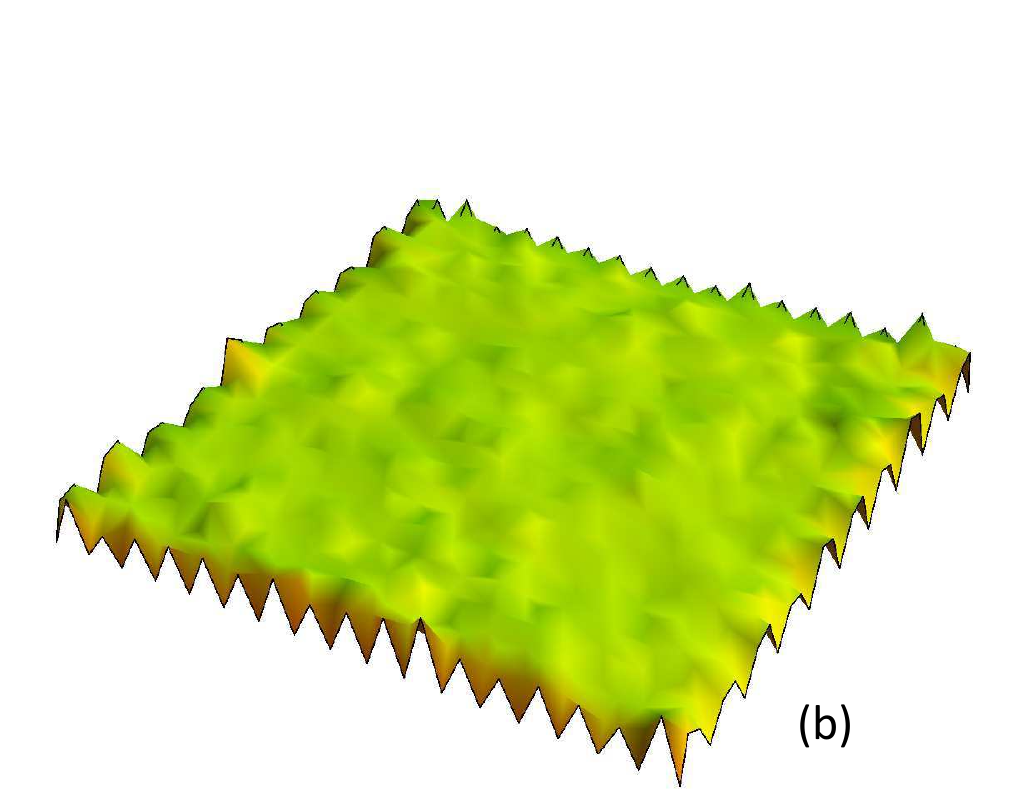}
\caption{(Color online) (a) The disorder averaged edge localization 
$<P^{Edge}_{n}>$ vs. the disorder amplitude $W$ for several helical states; 
the states prove to be  robust against disorder only for small disorder 
($\Phi=0$, $\lambda_{SO}=0.05$, no. of lattice sites is $35 \times 20$, 
no. of disorder configurations is 880).
(b) The  disorder averaged wave function $<|\Psi(r)|^2>$
for a former helical state at $\Phi=0$
($\lambda_{SO}=0.05$, $W=5$, no. of lattice sites
is $35 \times 20$, no. of
disorder configurations is $500$).} 
\end{figure}

The level spacing analysis helps also to understand the disorder effects 
on the energy spectrum. Let us define the level spacing  as
$ t_n=\delta E_n/<\delta E_n>$, 
where $\delta E_n=E_{n+1}-E_{n}$ and  $<...>$ means the average over 
all disorder configurations.
The level spacing distributions $P_n(t)$ calculated numerically
at low disorder for several $n$-s corresponding to states in the 
topological gap are shown in Fig.7b. The distributions can be well fitted with 
Gaussian functions. Since the same distribution is exhibited also by the 
edge states in the integer quantum Hall phase \cite{Nita1}, one may
conclude that the Gaussian distribution of the level spacing is the
feature of the states that are robust against disorder. 
It is to note in Fig.7b that the curves 
show different widths,  namely the  states in the center of the topological gap
exhibit a narrower width (being more robust) than those located near 
the gap margin.

Additional information can be obtained by calculating the edge localization 
Eq.(6) as function of $W$. 
In Fig.8a we find that the disordered helical states  remain  localized 
near  edges as long as $W$ is small. However, $P^{Edge}$ falls down
with increasing disorder, meaning that the 
states extend gradually 
inside the plaquette, and eventually become disordered metallic-like 
states spread over the whole plaquette area if $W \gtrsim 4 $.
An example of such a uniformly distributed state 
originating from a helical state is shown in Fig.8b. Of course, the level
spacing distribution should change also from the Gaussian to a Wigner-
Dyson distribution, however this topic will be discussed elsewhere. 

\section{Specific IQHE and QSHE of the confined grahene with spin-orbit
coupling }

In this section, we  simulate a four-lead device by attaching leads to a
graphene plaquette, and  calculate the longitudinal and transverse
resistances corresponding to both spin and charge currents. 
We emphasize  specific properties of the transmittance matrix 
$T_{\alpha\beta}^{\sigma}$ in the presence of the ISO coupling that 
generate a uncommon behavior of the spin and charge quantum Hall effect.
We find that, besides the usual plateaus of the Hall conductance at  
$\pm (2 e^2/h)(2n+1)$, the IQHE gets new intermediate plateaus at 
$\pm (2e^2/h)(2n+2)$ with  $n=0,1,2,...~ $.
Next, we find  that these plateaus  are  associated with a non-vanishing  
quantum spin conductance $G_H^S=- 2e/4\pi$, the sign being opposite
to the usual spin Hall conductance that occurs in the topological gap. 
The changes of both IQHE and QSHE can be observed in Fig.12, which
represents the main result of the section. These transport effects have 
not been  explored up till now, and it turns out that they  shows up 
in  energy gaps with  spin imbalance,  where 
$T_{\alpha,\alpha+1}^{\uparrow} \ne T_{\alpha,\alpha+1}^{\downarrow}$.
The transport calculations are based on the Landauer-B\"uttiker
formalism.

\subsection{Properties of the spin-resolved transmittances}
The  Landauer-B\"uttiker approach  requires that the Hamiltonian (1)  
be completed with terms describing the leads ($H_L$) and the coupling 
between leads and the graphene plaquette ($H_{LP}$).
Then the total Hamiltonian reads:
\begin{equation}
H^T= H+H_L+\tau H_{LP},
\end{equation}
where the last two terms are considered to be spin independent.
The quantities which enter the expression of the spin-dependent electron 
transmittance between the leads $\alpha$ and $\beta$ are the lead-plaquette coupling $\tau$, the matrix element of the retarded Green function 
corresponding to the total Hamiltonian Eq.(7) and the lead density of 
states:
\begin{equation}
T^{\sigma}_{\alpha\beta}(E,\Phi)=4\tau^4 |G^{\sigma}_{\alpha \beta}|^2 
(E,\Phi) Im g^L_{\alpha}(E) Im g^L_{\beta}(E),~~\alpha \ne \beta~,
\end{equation}
where $g^L$ is the Green function of leads. The symmetries of the energy 
spectrum,  discussed in the previous section,  determine the properties 
of the Green function $G^{\sigma}_{\alpha \beta}$, and reflect eventually,
via Eq.(8), in the  symmetries of the spin-resolved transmittance matrix:
\begin{equation}
T_{\alpha\beta}^{\sigma}(E,\Phi)=T_{\beta\alpha}^{-\sigma}(-E,\Phi)
=T_{\beta\alpha}^{-\sigma}(E,-\Phi).
\end{equation}

An inspection of the Hofstadter spectrum shows that the  gaps located 
in the central part of the spectrum (i.e., topological and relativistic 
gaps) open at  lower magnetic fluxes  in comparison with the 
conventional Landau gaps located at higher energies.
That is, it is hard to find a magnetic flux that allows  to evidentiate 
simultaneously all the types of edge states occuring in different  
energy gaps. Intending to compare all the three regimes (topological, 
relativistic and conventional Landau), we need to perform the calculations
at a relatively small flux, and we select $\Phi=0.03\Phi_0$. 
On the other hand, at such a small flux,  the transmittance 
$T_{\alpha,\alpha+2}$  (that hops over a lead) gets rather large values 
in some energy ranges affecting the quantum Hall effect; the implications  
will be discussed below.

\begin{figure}[htb]
\includegraphics[angle=-00,width=0.5\textwidth]{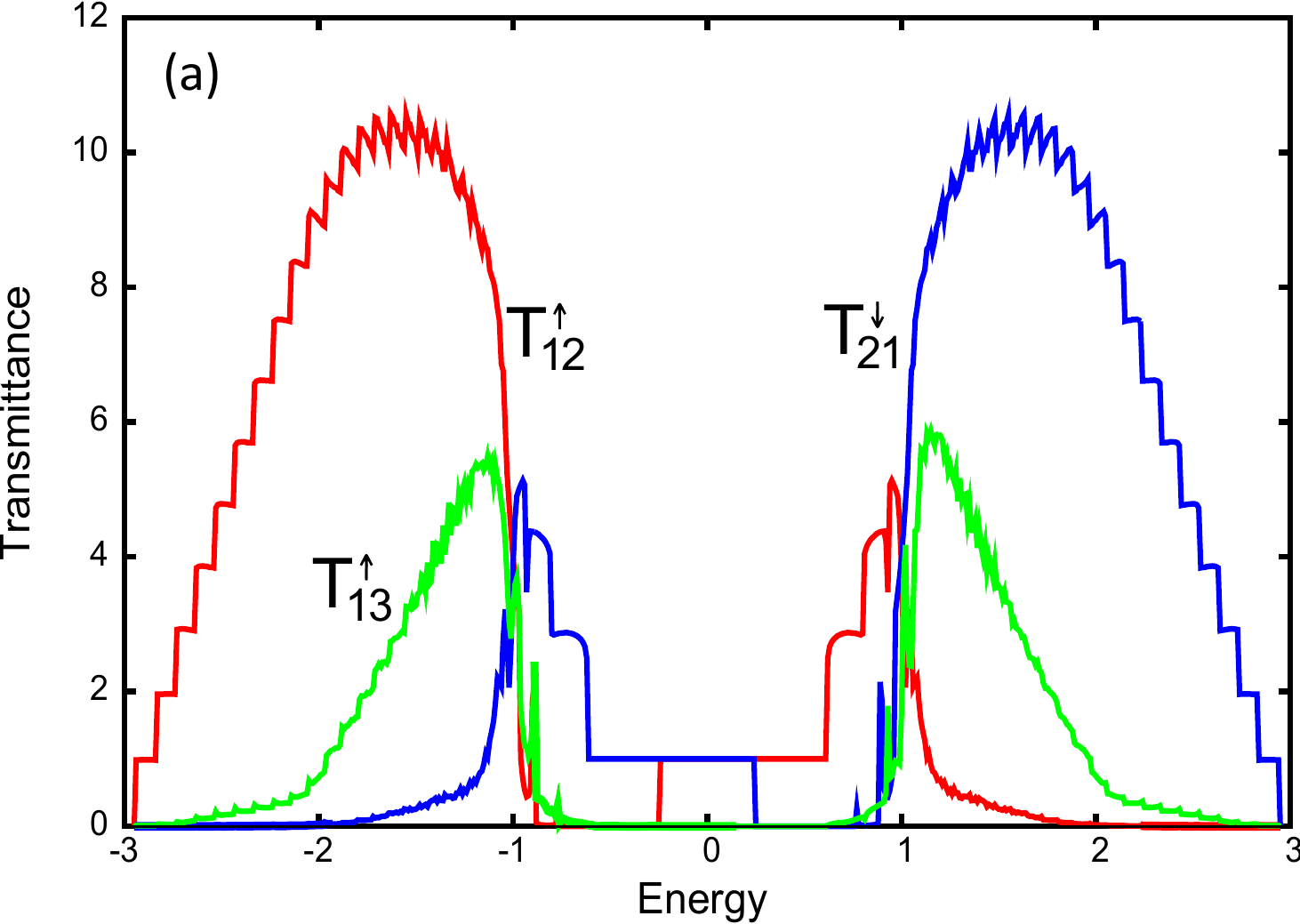}
\includegraphics[angle=-00,width=0.5\textwidth]{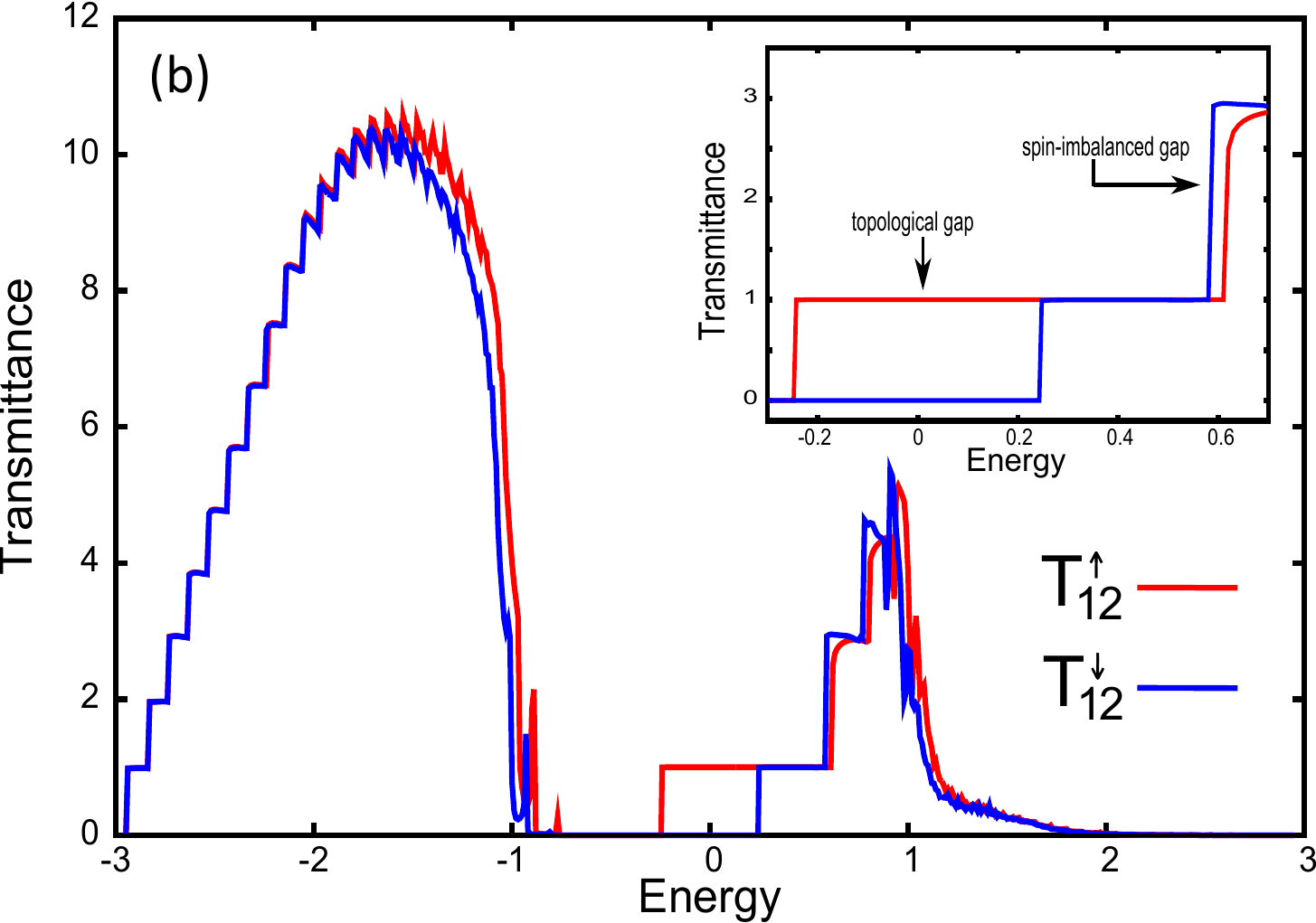}
\caption{(Color online) (a) Illustration of the symmetry 
$T_{12}^{\uparrow}(E) = T_{21}^{\downarrow}(-E)$; the transmittances  
show plateaus corresponding to all gaps present in the spectrum. 
$T_{13}$ is symmetric around $E=0$ and vanishes in the topological 
and relativistic regions, but it is not negligible in the rest of the 
spectrum. (b) $T_{12}^{\uparrow}$ and $T_{12}^{\downarrow}$ coincide 
at high energies, but are shifted in the middle of the spectrum. 
The inset shows the topological gap, where
$T_{12}^{\uparrow}=1$, $T_{12}^{\downarrow}=0$, and the imbalanced-gap,
where $T_{12}^{\uparrow}=1$, $T_{12}^{\downarrow}=3$  
($\Phi=0.03\Phi_0$, $\lambda_{SO}=0.05$, number of sites is $105 \times 40$).}
\end{figure}

The transmittances calculated according to Eq.(8) are illustrated in Fig.9.
In Fig.9a, the plot of  $T_{12}^{\uparrow}$ and $T_{21}^{\downarrow}$  
prove the symmetry $T_{12}^{\uparrow}(E)=T_{21}^{\downarrow}(-E)$ 
expressed by Eq.(9). The two transmittances allow an easy  identification 
of the topological gap in the middle of the energy-axis  
(approximately in the range $E\in[-0.25,0.25]$),  
where $T_{12}^{\uparrow}=T_{21}^{\downarrow}=1$,
denoting the presence of two channels of opposite spin running in
opposite directions, i.e., the well-known condition for QSHE.

In Fig.9b we compare the transmittance $T_{12}^{\uparrow}$ with 
$T_{12}^{\downarrow}$, which carry the opposite spin but runs in the 
same direction. 
The two curves coincides at high energy, however they show a 
significant shift in the central region of the spectrum. 
This  shift of the transmittances is an obvious consequence of 
the shifted spin-dependent energies in the spectrum,
which has been already noticed in Fig.5.
The inset depicts the energy range that comprises the topological 
gap and the spin-imbalanced gap about $E=0.6$, where we have to realize 
that $T_{12}^{\uparrow}=1$ and $T_{12}^{\downarrow}=3$. As the other 
transmittances vanish, we may conclude that in this gap there are 
four active channels, one of spin-up and three of spin-down, 
all of them running in the same direction. In the next  subsection, 
when calculating the transverse resistance, we shall see  that this 
spin-imbalance yields unusual plateaus of both integer and spin  Hall 
effect.

The relatively small value of the magnetic flux, at which we are  
compelled to perform the transport calculations, makes it interesting 
to discuss the behavior of $T_{\alpha,\alpha+2}^{\sigma}$. 
We remind  that at strong  magnetic fields, the gaps corresponding to 
the quantized plateaus are characterized by 
$T_{\alpha,\alpha+1}^{\sigma}$=integer, while
all the other transmittances vanish,
including $T_{\alpha,\alpha+2}^{\sigma}$.
At such  strong  fields, the edge states are localized very close 
to the perimeter of the plaquette, and the negligible value of  
$T_{\alpha,\alpha+2}$ may  be considered as a measure of the high degree 
of edge localization.
For $\Phi=0.03\Phi_0$ the transmittance $T_{13}^{\uparrow}$  is shown in
Fig.9a.  One notices the vanishing of $T_{13}$ in the topological gap,
fact that  attests the strong localization of the helical edge states.
On the other hand, the large values shown in a rather wide range 
about $E=\pm 1$ can be associated with the bulky character of the 
quantum  states, accompanied by the absence of  quantum plateaus in the
corresponding energy range.

\subsection{Longitudinal and transverse charge and spin conductances}
Without spin-flip processes, the system behaves 
as a two independent spin fluids. Then, the particle current in  
a multi-lead device can be written in the linear approach as
$I_{\alpha}^{\sigma}= \sum_{\beta\sigma}T_{\alpha\beta}^{\sigma}
V_{\beta}^{\sigma}$, where $\{\alpha,\beta\}$ stand for the lead indices,
$I_{\alpha}^{\sigma}$ is the current through the lead $\alpha$
and $V_{\beta}^{\sigma}$ is the potential at the contact site $\beta$. 
Summing up the contributions of the two spin, the total 
charge- and spin-currents flowing through the lead $\alpha$ read:
\begin{equation}
I_{\alpha}^Q=\frac{e^2}{h}\sum_{\sigma}I_{\alpha}^{\sigma},~~
I_{\alpha}^S =\frac{e}{4\pi}\sum_{\sigma}\sigma I_{\alpha}^{\sigma}.
\end{equation}
Since the transmitance matrix $T_{\alpha,\beta}^{\sigma}$ is already known,
the  spin-resolved   longitudinal and Hall  resistances can be calculated 
according to the Landauer-B\"{u}ttiker formalism as:
\begin{eqnarray}
\nonumber R_L^{\sigma}&=&R_{14,23}^{\sigma}=(T_{24}^{\sigma}T_{31}^{\sigma}
-T_{21}^{\sigma}T_{34}^{\sigma})/D, \\
R_H^{\sigma}&=&(R_{13,24}^{\sigma}-R_{24,13}^{\sigma})/2=
(T_{23}^{\sigma}T_{41}^{\sigma}-T_{21}^{\sigma}T_{43}^{\sigma}-
T_{32}^{\sigma}T_{14}^{\sigma}+T_{12}^{\sigma}T_{34}^{\sigma})/2D,
\end{eqnarray}
where $D$ is any 3x3  subdeterminant of the transmittance matrix.

The comparative study of the Hall resistance, with and without SO coupling,
is  instructive. This is done in Fig.10a,b where  $R_H^{\sigma}$, 
calculated at the flux  $\Phi=0.03$, is superimposed over the 
corresponding Hofstadter spectrum.
When the spectrum and the resistance are spin independent 
(the case $\lambda=0$), the two quantities looks like in Fig.10a, 
where the antisymmetry  $R_H^{\sigma}(E,\Phi)=-R_H^{\sigma}(-E,\Phi)$ 
is obvious. If $\lambda\ne 0$, the  previous antisymmetry at the 
reflection $E\rightarrow -E$  is lost since  one has now to inverse  
also the spin. 
Indeed, from Eq.(9) and Eq.(11), it follows immediately that 
$ R_H^{\sigma}(E,\Phi)=-R_H^{-\sigma}(-E,\Phi)$.

\begin{figure}[htb]
\includegraphics[angle=-00,width=0.5\textwidth]{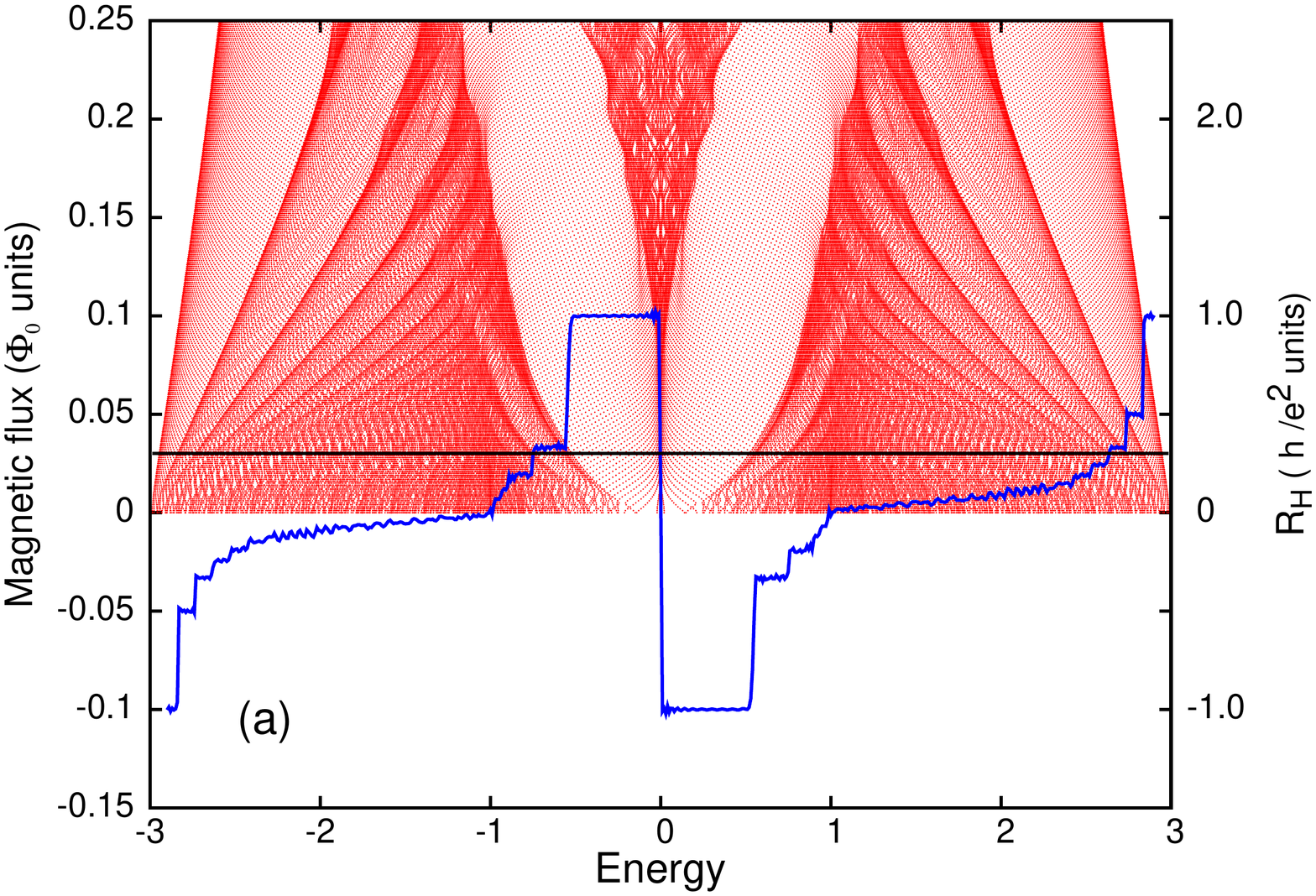}
\includegraphics[angle=-00,width=0.5\textwidth]{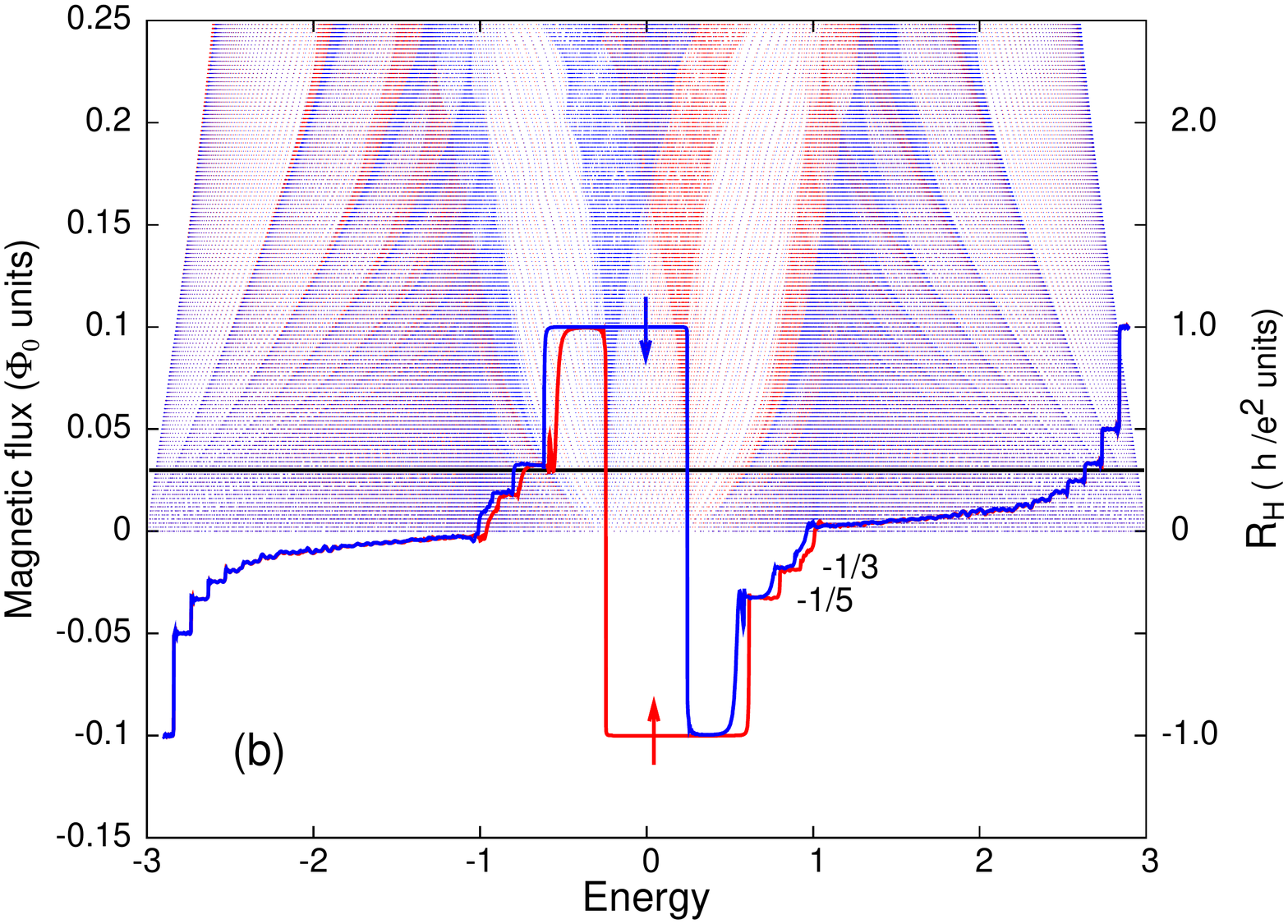}
\caption{(Color online) The spin-resolved Hall resistances $R_H^{\sigma}$
at $\Phi=0.03$ (the flux is indicated by the black horizontal line): 
(a) at $\lambda_{SO}=0$, the two resistances for spin up and down 
coincide and are symmetric around $E=0$. 
(b) at $\lambda=0.05$, $R_H^{\uparrow}$ (red curve) and  
$R_H^{\downarrow}$ (blue curve) are different,  and   satisfies the 
property $R_H^{\uparrow}(E)=-R_H^{\downarrow}(-E)$ mentioned in the text;
the difference is more visible in the central region $E\in (-1,1)$.
Notice that the plateaus at $\pm e^2/h$  cover the topological and  
first relativistic  gap, confirming the behaviour of edge states 
sketched in Fig.6.  The plaquette consists of $33\times 30$ sites.}
\end{figure}

It is to emphasize that in Fig.10b the spin-dependent Hall resistances
shows the plateaus $R_H^{\downarrow}=1$ and $R_H^{\uparrow}=-1$,
respectively, which extend over both the topological gap and the 
the first relativistic gap, each one crossing the band of opposite spin. 
This behavior of the resistance is the immediate result of the 
fact that the chiral edge states located in the first relativistic gap
extend also in the topological gap, as underlined in the previous section
and observable in Fig.3(right). 

Since the two spins act as two parallel channels, the total charge and spin
resistances ($R^Q$ and $R^S$, respectively) are given by:
\begin{equation}
\frac{1}{R_{L,H}^Q}=\frac{1}{R_{L,H}^{\downarrow}}+
\frac{1}{R_{L,H}^{\uparrow}},~~
\frac{1}{R_{L,H}^S}=\frac{1}{R_{L,H}^{\downarrow}}-
\frac{1}{R_{L,H}^{\uparrow}},
\end{equation}
where the indices $L$ and $H$  stand for longitudinal and
transversal (Hall), respectively.
It is straightforward to prove the relationships:
\begin{equation}
R_{H}^Q(E)=-R_{H}^Q(-E), ~~
R_{H}^S(E)=R_{H}^S(-E),
\end{equation}
that show the different symmetries  of the charge and spin resistance.

In the previous section, we mentioned the presence in the relativistic 
range of the energy spectrum of small gaps, induced by the spin-orbit 
splitting, where a spin imbalance exists. Obviously, the imbalance  
gives rise  to $R_H^{\uparrow}\ne R_H^{\downarrow}$, and, 
in line with Eq.(12), this fact indicates the presence of a  net 
spin current.
In other words, the mesoscopic graphene plaquette exhibits QSHE
not only in the weak topological gap, located symmetrically about $E=0$,
but also  in some other energy stripes, where   the number
of spin-up edge states differs from the spin-down edge states.
Another peculiar aspect is that, contrary to the situation in the 
topological gap, the two spins  flow in the same direction,
as the chirality of the edge states is the same no matter the spin. 

\begin{figure}[htb]
\hskip-2cm
        \includegraphics[angle=-00,width=0.55\textwidth]{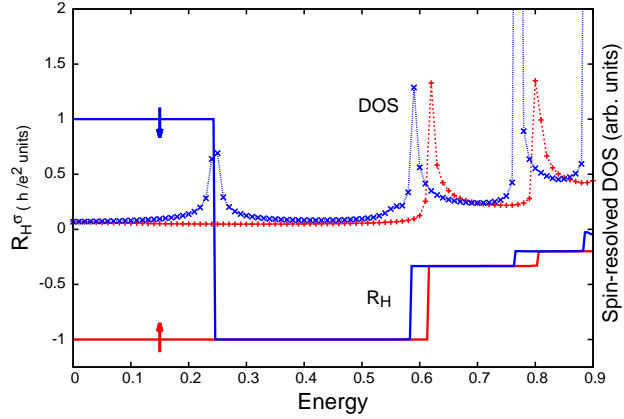}
\caption{(Color on line) The spin resolved Hall resistance and density 
of states; the shifted regions are those where the spin resistance 
should be non-zero. The densities of states exhibit the same shift 
($\Phi=0.03$, $\lambda_{SO}=0.05$, number of sites= $105 \times 40$).}
\end{figure}

The result of the numerical calculation for
$R_H^{\uparrow}$ and $R_H^{\downarrow}$ is presented in Fig.11 together 
with the spin-dependent densities of states \cite{note5}.
The shift between the spin-up and spin-down resistances, visible
in the energy stripes about $E\sim 0.6$ and $E\sim 0.8$, is confirmed 
by a similar shift between the two spin-dependent densities of states.

A last comment concerns the sign of the QSHE: in the topological gap
one has $R_H^{\downarrow}=1$ and $R_H^{\uparrow}=-1$; then, according to 
Eq.(12), the sign of the total Hall resistance $R_H^S$ is positive. 
On the other hand, in the spin-imbalanced gap at $E=0.6$, one has  
$R_H^{\downarrow}=-1/3$, and $R_H^{\uparrow}=-1$, 
meaning that $R_H^S$ is negative.  This opposite sign of $R_H^S$ follows from the 
fact that, in the spin-imbalanced gap, the  spin-up 
and spin-down edge sates show the  same chirality 
(which is not the case in the topological gap).

\begin{figure}[htb]
        \includegraphics[angle=-0,width=0.6\textwidth]{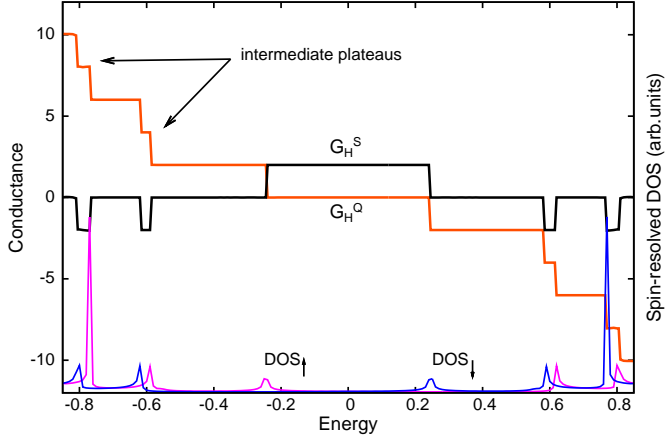}
\caption{(Color online) The spin and charge Hall conductance in the 
quantum regime (in $2e/4\pi$ and $e^2/h$ units, respectively):  
novel plateaus are visible in the imbalanced gaps 
opened by the intrinsic spin-orbit coupling. The spin-resolved
density of states are also shown ($\Phi=0.03$,
$\lambda_{SO}=0.05$, number of sites= $105 \times 40$).}
\end{figure}

The longitudinal and Hall resistances  being known from Eq.(12), 
in the spirit of the experimental work \cite{Novoselov}, we shall plot 
in Fig.12 the corresponding Hall conductances calculated as:
\begin{equation}
G_H^Q=\frac{R_H^Q}{(R_H^Q)^2+(R_L^Q)^2},~~
G_H^S=\frac{R_H^S}{(R_H^S)^2+(R_L^S)^2}.
\end{equation}
In what concerns the charge conductance $G_H^Q$, one has to observe 
not only the vanishing value in the topological range and the known 
plateaus at 2, 6 and 10  in the relativistic one, 
but also some unexpected plateaus  at 4 and 8 (in units $e^2/h$).
A similar behavior is proved by the spin Hall conductance $G_H^S$,
that shows the expected value  2$e/4\pi$ in the topological range, 
and then vanishes everywhere except the same energy stripes where 
the unusual values of the charge Hall conductance occur. 
In the respective stripes the spin Hall conductance equals $-2e/4\pi$.
According to the previous discussions, it is obvious that they appear 
in the spin-imbalanced gaps generated by the intrinsic spin-orbit 
interaction in the presence of the magnetic field.

\section{Conclusions}
The main result consists in finding anomalous plateaus 
$G_H^S=-2e/4\pi$ of the QSHE outside the topological gap, namely in 
the range of the {\it spin-imbalanced} gaps.
In the same places, the IQHE exhibits also uncommon intermediate plateaus
at $G_H^Q=\pm(4e^2/h)(n+1)$. The spin-imbalanced gaps are characterized by 
a non-equal number of spin-up and spin-down edge states. 
They are due to the splitting generated by ISO coupling, as sketched in Fig.6.
Since we consider a small spin-orbit coupling ($\lambda_{SO}<< t$), 
this type of gaps appear in the relativistic range of the energy spectrum.

In the 'weak' topological gap, the degeneracy of the helical states is lifted 
by the magnetic field $B$. The states evolve with $B$ and merge into  the bands
that confine the  gap. During this process the states lose
the edge character, as shown in Fig.4. 
At some higher magnetic fields, the topological gap gets filled with 
edge states of
chiral origin, i.e., coming from the neighboring Dirac-Landau gaps. 
 These states come also in pairs of opposite spin and chirality 
(see Fig.3(right)), 
so that the QSHE survives, and
 continues to equal $2e/4\pi$, even at higher magnetic fields,
as long as the gap remains open.

We have noticed that, outside the  topological gap, in the relativistic gaps
of the energy spectrum, the degree of  localization of the edge states 
depends on the spin. Moreover, the derivatives $dE^{\uparrow}/d\Phi$
and $dE^{\downarrow}/d\Phi$ are different, meaning that also the diamagnetic
moments carried by states of opposite spins differ in magnitude.

We have looked for  disorder effects  on the topological gap and helical
edge states at vanishing magnetic field. Although being robust at 
low disorder, the localization along the edges of the helical states 
is lost at higher disorder strength (see Fig.8a), 
when the states become of metallic-type, being distributed uniformly on the
whole plaquette. 
We find also that   the level spacing of the disordered helical states 
follow a Gaussian distribution at low disorder (see Fig.7b).
For $W\gtrsim 2$, the topological gap at $\Phi=0$ 
becomes progressively narrower with increasing disorder
under the compression of the levels that stem from the highly 
quasi-degenerated regions about $E=\pm 1$,
resulting in a {\it tulip}-like spectrum (as in  Fig.7a).

The transmittance matrix is spin dependent and its symmetries are shown in 
Eq.(9) and Fig.9a. The symmetry of $T_{\alpha \beta}^{\sigma}(E,\Phi)$
results in the properties of the spin and charge Hall resistance of the
four-lead graphene device, which we are interested in. The numerical
calculations are performed at relatively small magnetic flux 
($\Phi=0.03\Phi_0$) in order to catch the effect of all gaps specific to
graphene (i.e., of Dirac-Landau and conventional Landau type). 
This  allows also to show  the energy dependence of 
$T_{\alpha \alpha+2}^{\sigma}(E,\Phi)$, the transmittance that kills 
the quantum plateaus when it takes large values (noticeable in Fig.9a).

\section{Acknowledgements}
We acknowledge  support from PNII-ID-PCE Research Programme
(grant no. 0091/2011) and Core Programme (contract no. 45N/2009).


\begin{references}
\bibitem{Novoselov} K. S. Novoselov, A. K. Geim, S. V. Morozov, D. Jiang, 
M. I. Katsnelson, I. V. Grigorieva, S. V. Dubonos, and  A. A. Firsov,
Nature {\bf 438}, 197 (2005).

\bibitem{Kane} C. L. Kane and E. J. Mele, Phys. Rev. Lett. {\bf 95},
226801 (2005).

\bibitem{Kane-Z2} C. L. Kane and E. J. Mele, Phys. Rev. Lett. 
{\bf 95}, 146802 (2005).

\bibitem{Volkov} B. A. Volkov and A. O. Pankratov,
Pis'ma Zh. Eksp. Teor. Fiz. {\bf 42}, 145 (1985).

\bibitem{Bernevig} B. A. Bernevig, T. L. Hughes, S. C. Zhang, 
Science {\bf 314}, 1757 (2006).

\bibitem{Koenig}  M. K\"{o}nig, S. Wiedmann, C. Br\"{u}ne, A. Roth,
H. Buhmann, L. W. Molenkamp, X. L. Qi, S. C. Zhang,
Science {\bf 318}, 766 (2007).

\bibitem{Liu} C. Liu, T. L. Hughes, X.L. Qi, K. Wang, and S.C. Zhang, 
Phys. Rev. Lett. {\bf 100}, 236601 (2008).

\bibitem{Knez}L. Du, I. Knez, G. Sullivan, R. R. Du,
arXiv:1306.1925 [cond-mat.mes-hall].

\bibitem{Hasan} M. Z. Hasan and M. Z. Kane, 
Rev. Mod. Phys. {\bf 82}, 3045 (2010).

\bibitem{Ando} Y. Ando, J. of Phys. Soc. Japan {\bf 82}, 102001 (2013).

\bibitem{Gibertini} M. Gibertini, A. Singha, V. Pellegrini, M. Polini,
 G. Vignale, A. Pinczuk, L. N. Pfeiffer, and K. W. West, 
Phys. Rev. B 79, 241406(R) (2009).

\bibitem{Tarruell} L. Tarruell, D. Greif, T. Uehlinger, G. Jotzu, 
and T. Esslinger, Nature  {\bf 483}, 302 (2012).

\bibitem{Weeks} C. Weeks, J. Hu, J. Alicea, M. Franz, and R. Wu,
Phys. Rev. X {\bf 1}, 021001 (2011).

\bibitem{Balakrishnan} J. Balakrishnan, G. K. W. Koon, M. Jaiswal,
A. H. Castro Neto, and  B. \"{O}zyilmaz,  Nature Physics {\bf 9},
284 (2013).

\bibitem{Ezawa}  M. Ezawa, J. Phys. Soc. of Jpn. {\bf 81}, 064705, (2012). 

\bibitem{Liu1} C. C. Liu, W. Feng, and Y. Yao, Phys. Rev. Lett. {\bf 107}, 
076802 (2011).

\bibitem{Brey} L. Brey, H. A. Fertig, Phys. Rev. B {\bf 73}, 235411 (2006).

\bibitem{Montambaux} P. Delplace, D. Ullmo, G. Montambaux,
Phys. Rev. B {\bf 84}, 195452 (2011).

\bibitem{Shevtsov} O. Shevtsov, P. Carmier, C. Petitjean, C. Groth,
D. Carpentier, and X. Waintal,
Phys. Rev. X {\bf 2}, 031004 (2012).

\bibitem{Beugeling} W. Beugeling, N. Goldman, and C. M. Smith, 
Phys. Rev. B {\bf 86}, 075118 (2012).

\bibitem{note} In the continuous model for the graphene ribbon 
\cite{Kane-Z2} the width of the gap 
at  $B=0$ is  $\Delta=6\sqrt 3\lambda_{SO}$, where $\lambda_{SO}$ 
is the spin-orbit coupling constant.

\bibitem{note1}
Meaning $E_a\ne E_b$ in Hamiltonian (2). 

\bibitem{Rammal} R. Rammal, J. Phys. France {\bf 46}, 1345 (1985).

\bibitem{Cuniberti} N. Nemec and G. Cuniberti Phys. Rev. B {\bf 74}, 165411 (2006).

\bibitem{note2} In the continuous model, the relativistic bands depend on 
the magnetic field as $\sqrt B$, while the conventional Landau bands 
are linear. For the finite plaquette described by the tight-binding model,
this remains true at low magnetic flux, as it can be seen  in Fig.3(left).
The differences are due to the discrete lattice structure and to
the confinement.

\bibitem{Nita} M. Ni\c t\u a, B. Ostahie, and A. Aldea, Phys. Rev. B {\bf 87},
125428 (2013).

\bibitem{note3} 
Another way to create  such an imbalance is by  imposing staggered 
energies of atoms A and B in the honeycomb lattice \cite{Beugeling}.

\bibitem{Peres}
N. M. R. Peres, F. Guinea, and A. H. Castro Neto, Phys. Rev. B {\bf 73}, 
125411 (2006).

\bibitem{note4}
Another way to 
consider the spin-imbalance is by noting that 
the spin-orbit gap is an intersection of the first spin-up 
relativistic gap with the second spin-down relativistic gap. 
Then one has obviously $N_{\uparrow}=1$ and $N_{\downarrow}=3$, 
i.e., the number of spin-up and spin-down edge states are different 
in the spectral range corresponding to the intersection.

\bibitem{NW} J. von Neumann and E. Wigner, Phys. Z. 30, 467 (1926).

\bibitem{Demkov} Yu. N. Demkov, P. B. Kurasov, Teoret. Mat. Fiz. {\bf 153},
 68 (2007).

\bibitem{Nita1}
M. Ni\c t\u a,  A. Aldea, and  J. Zittartz, 
Phys. Rev. B {\bf 62}, 15367 (2000).

\bibitem{note5}
The spin dependent DOS is defined as 
$-(1/\pi)Im \sum_i G_{ii}^{\sigma}(E+i0)$.

\end{references}
\end{document}